\documentclass[preprint,superscriptaddress,showpacs,10pt]{revtex4-1}
\usepackage{amssymb}
\usepackage{amsmath}
\usepackage{hyperref}

\begin{document}
\title{The Entropy Relations of Black Holes with Multihorizons in Higher Dimensions}
\author{Jia Wang}
\email{wangjia2010@mail.nankai.edu.cn}
\affiliation{School of Physics, Nankai University, Tianjin 300071, China}
\author{Wei Xu}
\email{xuweifuture@mail.nankai.edu.cn}
\affiliation{School of Physics, Nankai University, Tianjin 300071, China}
\author{Xin-he Meng}
\email{xhm@nankai.edu.cn}
\affiliation{School of Physics, Nankai University, Tianjin 300071, China}
\affiliation{State Key Laboratory of Theoretical Physics, Institute of Theoretical Physics, Chinese Academy of
             Sciences, Beijing 100190, China}
\affiliation{Kavli Institute of Theoretical Physics China, Chinese Academy of Sciences, Beijing 100190, China}

\begin{abstract}
We study the entropy relations of multi-horizons black holes in higher dimensional (A)dS spacetime with maximal symmetries, including Einstein-Maxwell gravity and $f(R)$(-Maxwell) gravity. These additional equalities in thermodynamics are expected to be useful to understanding the origin of black hole entropy at the microscopic level. Revisiting the entropy product introduced by Cvetic etc, in our case, it has an unexpected behavior. It is shown that this electric charge $Q$ plays an important role in this entropy product. The entropy product of charged black holes only depends on the electric charge $Q$ and is mass independence. When $Q$ vanishes in the solution, it turns to mass dependence, even when including the effect of the un-physical ``virtual'' horizons. In this sense, the ``universal relation'' of this entropy product is destroyed. Then we introduce another kind of ``universal'' entropy relation, which only depends on the cosmological constant $\Lambda$ and the background topology $k$, and which does not depend on the conserved charges $Q$, nor even the mass $M$.
\end{abstract}
\maketitle

\section{Introduction}
Since the thermodynamics of black holes has been established, the entropy of horizon is a fascinated research
interest. The recent popular topic is the additional equalities in thermodynamics, which are expected to be useful of understanding the origin of black hole entropy at the microscopic level. These equalities include the entropy product of multi-horizons black hole in super-gravity model \cite{Cvetic:2010mn,Toldo:2012ec,Cvetic:2013eda,Lu:2013ura,Chow:2013tia}, Einstein gravity \cite{Detournay:2012ug,Castro:2012av,Visser:2012zi,Chen:2012mh,Castro:2013kea,Visser:2012wu,Abdolrahimi:2013cza,Pradhan:2013hqa} and other modified gravity models \cite{Castro:2013pqa,Cvetic:2013eda,Faraoni:2012je,Lu:2013eoa,Anacleto:2013esa} in both four and high dimensions, and the entropy sum \cite{Wang:2013smb,Xu:2013zpa} of multi-horizons black hole. It is always shown that the entropy product and sum are often independent of the mass of the black hole \cite{Cvetic:2010mn,Castro:2012av,Toldo:2012ec,Chen:2012mh,Visser:2012zi,Cvetic:2013eda,Abdolrahimi:2013cza,Lu:2013ura,
Anacleto:2013esa,Chow:2013tia,Castro:2013kea,Lu:2013eoa,Wang:2013smb,Xu:2013zpa}, while the former fails in some asymptotical non-flat spacetime \cite{Faraoni:2012je,Castro:2013pqa,Detournay:2012ug,Visser:2012wu}. However, in order to preserve the mass independence, one need include the necessary effect of the un-physical ``virtual'' horizons \cite{Visser:2012wu,Wang:2013smb,Xu:2013zpa}. Only in this way, these additional equalities of multi-horizons of black holes are ''universal''. On the other hand, by using of the new thermodynamics relations the construction of thermodynamics for inner horizon of black hole catches more attentions recently \cite{Detournay:2012ug,Castro:2012av,Chen:2012mh,Pradhan:2013hqa,Castro:2013pqa,Pradhan:2013xha,Ansorg:2008bv,Ansorg:2009yi,Ansorg:2010ru}. These make investigating the properties of inner horizons of black hole more interesting, especially for the above additional equalities of multi-horizons to be found.

After looking at the entropy product or sum of multi-horizons of the black hole, one can find that they only depend on the conserved charges: the electric charge $Q$, the angular momentum $J$, or the cosmological constant $\Lambda$ (which can be treated as pressure in AdS spacetime after  explaining the mass of the black hole as  enthalpy rather than internal energy of the system). For example, the area product (entropy product) of stationary $(3+1)$ dimensional black hole is $Q$ and $J$ dependence \cite{Ansorg:2008bv,Ansorg:2009yi,Ansorg:2010ru}
\begin{align}
A_{C}A_{E}=(8\pi)^2\left(J^2+\frac{Q^4}{4}\right).
\end{align}
where $A_{C}$ and $A_{E}$ denote the horizon areas of inner Cauchy horizon and event horizon
respectively. The conclusion is generalized to rotating multicharge black holes, both in asymptotically flat and
asymptotically anti-de Sitter spacetime in four and higher dimensions \cite{Cvetic:2010mn}. The $\Lambda$ dependence is shown in the entropy sum of four dimensional RN-(A)dS black hole \cite{Wang:2013smb,Xu:2013zpa}: $\sum S_i=\frac{6\pi k}{\Lambda}.$ In this paper, we introduce more types of entropy products
and consider their conserved charge dependence in high dimensions. We are aimed in high dimensions, as studies of gravity in higher dimensions might reveal deeper structures of general relativity, such as stability issues and classification of singularities of spacetime in various dimensions. Entropy products of multihorizons in four dimensions help us to understand the origin of black hole entropy at the microscopic level. In this sense,  the entropy relations in high dimensions might also lead to a further studying on this issue. However,  this is not a new idea, as in \cite{Visser:2012wu}, some interesting entropy product with mass independence were shown in four dimensions. Generalizations to higher dimensional spacetime, the results in this paper agree with this aim and have shown a whole look at the extra thermodynamical relationships, especially for some ``universal'' and mass independent entropy product. Firstly we find this type of entropy product, $\sum_{1\leq i<j\leq D}(S_{i}S_{j})^{\frac{1}{d-2}}$, only depends on the cosmological constant $\Lambda$ and horizon topology $k$, and which does not depend on the other conserved charges. Here and below we use $D$ to denote the number of the roots of polynomial, i.e. the number of the horizons. For another type of charged black hole the $\prod_{i=1}^DS_i$ only depends on the electric charge $Q$. However, the electric charge $Q$ plays a switch role in latter case. When $Q$ vanishes in the solution, the entropy product turns to mass dependence.

This paper is organized as follows. In the next Section, we will investigate the``entropy product''
of higher dimensional (A)dS black holes in Einstein-Maxwell Gravity. In Section 3, we discuss the``entropy product'' of (A)dS black holes in $f(R)$-Maxwell Gravity. Section 4 is devoted to the conclusions and discussions.

\section{Higher Dimensional (A)dS Black Holes in Einstein-Maxwell Gravity}
\subsection{Charged Black Holes}
The Einstein-Maxwell Lagrangian in higher dimensions $d$ $(d\geq4)$ reads
\begin{align} \label{eq:Lag_of_Einstein-Maxwell}
  \mathcal{L}=\frac{1}{16\pi G}\int d^d x \sqrt{-g}[R-F_{\mu\nu}F^{\mu\nu}-2\Lambda],
\end{align}
where $\Lambda=\pm\frac{(d-1)(d-2)}{2\ell^2}$ is the cosmological constant associated with cosmological scale $\ell$. Here, the negative cosmological constant corresponds to AdS spacetime while positive one corresponds to dS spacetime. For a charged static black hole in maximal symmetric Einstein space, the metric takes the ansatz
\begin{align} \label{eq:metric ansatz}
  d s^2=-V(r) d t^2+\frac{d r^2}{V(r)}+r^2 d\Omega^2_{d-2},
\end{align}
where $d\Omega^2_{d-2}$ represents the line element of a
$(d-2)$-dimensional maximal symmetric Einstein space with constant
curvature $(d-2)(d-3)k$, and the $k=1, 0$ and $-1$, corresponding
to the spherical, Ricci flat and hyperbolic topology of the black
hole horizon, respectively. The metric function $V(r)$ is given by \cite{Liu:2003px,Astefanesei:2003gw,Brihaye:2008br,Belhaj:2012bg}
\begin{align}
  V(r)=k-\frac{2M}{r^{d-3}}+\frac{Q^2}{r^{2(d-3)}}-\frac{2\Lambda}{(d-1)(d-2)}r^2,
\label{V(r)1}
\end{align}
where $M$ is the mass parameter and $Q$ is the charge of the black hole. In the Einstein-Maxwell gravity, the entropy of horizon located in $r=r_i$, as usual, is given by (We have set $G=1$.)
\begin{align}
  S_i=\frac{A_i}{4}=\frac{\pi^{(d-1)/2}}{2\Gamma\left(\frac{d-1}{2}\right)}r_i^{d-2},
  \label{entropy1}
\end{align}
where $r_i$ which are the roots of $V(r)$, are the horizon coordinates. According to the horizon function (\ref{V(r)1}), in principle this high order polynomial has $D=2(d-2)$ roots. The real positive roots correspond to physical horizons while the negative or complex ones correspond to un-physical horizons which we mean ``virtual'' horizons. We can introduce some entropy relations, which are either conserved quantity independent or special conserved quantity dependent. Rewriting $V(r)=0$ by (\ref{V(r)1}), we obtain
\begin{equation}
\frac{2\Lambda}{(d-1)(d-2)}r^{2d-4}-kr^{2(d-3)}+2Mr^{d-3}-Q^2=0
\end{equation}
By using the Vieta theorem on the above equation, one can find
\begin{align}
&\sum_{1\leq i<j\leq D}r_{i}r_{j}=-\frac{k(d-1)(d-2)}{2\Lambda},\\
&\prod_{i=1}^Dr_i=-\frac{Q^2(d-1)(d-2)}{2\Lambda},\\
&\sum_{1\leq i_1< i_2\cdots< i_a\leq D}\,\prod_{a=1}^{d-1}r_{i_a}=(-1)^{d-1}\frac{M(d-1)(d-2)}{\Lambda}.
\end{align}
After inserting the inverse of the entropy (\ref{entropy1}) as
\begin{equation}
r_i=\left(\frac{2\Gamma\Bigl(\frac{d-1}{2}\Bigr)S_i}{\pi^{\frac{d-1}{2}}}\right)^{\frac{1}{d-2}}
\end{equation}
one can obtain the following type of entropy product
\begin{equation} \label{eq:varentropyproduct_of_E-M}
\sum_{1\leq i<j\leq D}(S_{i}S_{j})^{\frac{1}{d-2}}
    =-\frac{k(d-1)(d-2)\pi^{\frac{d-1}{d-2}}}{2\Lambda\left(2\Gamma\Bigl(\frac{d-1}{2}\Bigr)\right)^{\frac{2}{d-2}}}
\end{equation}
which only depends on the cosmological constant $\Lambda$ and horizon topology $k$. Another type of entropy product with the well-known mass independence is
\begin{equation}
\prod_{i=1}^DS_i=\left(-\frac{Q^2(d-1)(d-2)\pi^{d-1}}{8\Lambda\Gamma\Bigl(\frac{d-1}{2}\Bigr)^2}\right)^{d-2}
\end{equation}
Once again, it depends on cosmological constant $\Lambda$ and electric charge $Q$, except for mass $M$. The same phenomenon also happens in the entropy product study of (A)dS black holes \cite{Cvetic:2010mn,Visser:2012wu}. And the last one
\begin{align}
\sum_{1\leq i_1< i_2\cdots< i_a\leq D}\left(\prod_{a=1}^{d-1}S_{i_a}\right)
^{\frac{1}{d-2}}=(-1)^{d-1}\left(\frac{\pi^{\frac{d-1}{2}}}{2\Gamma\Bigl(\frac{d-1}{2}\Bigr)}\right)
                 ^{\frac{d-1}{d-2}}\frac{M(d-1)(d-2)}{\Lambda}.
\end{align}
is mass dependent.

\subsection{Neutral Black Holes}
If the Maxwell term vanishes in the Lagrangian (\ref{eq:Lag_of_Einstein-Maxwell}), the situation will be different from our discussion above. Then the Lagrangian is the standard Einstein case
\begin{align}
\mathcal{L}=\frac{1}{16\pi G}\int d^d x \sqrt{-g}(R-2\Lambda).
\end{align}
We still make the same metric ansatz (\ref{eq:metric ansatz}) i.e. the maximal symmetric, static black hole
solution. In this case the metric function reduces to
\begin{align}
V(r)=k-\frac{2M}{r^{d-3}}-\frac{2\Lambda}{(d-1)(d-2)}r^2.
\end{align}
In fact when $d=4$, the metric reduce to Schwarzschild (A)dS black holes.
Rewriting it to a polynomial equation of $r$, which shows the horizons of the black hole,
\begin{align}
\frac{2\Lambda}{(d-1)(d-2)}r^{d-1}-kr^{d-3}+2M=0.
\end{align}
The black hole possesses $D=d-1$ horizons at most, including ``virtual'' horizon. By the same help of Vieta theorem, we have
\begin{align}
&\sum_{1\leq i<j\leq D}r_{i}r_{j}=-\frac{k(d-1)(d-2)}{2\Lambda},\\
&\prod_{i=1}^Dr_i=(-1)^D\frac{M(d-1)(d-2)}{\Lambda}.
\end{align}
which immediately result in the following two entropy products
\begin{align}
&\sum_{1\leq i<j\leq D}(S_{i}S_{j})^{\frac{1}{d-2}}
    =-\frac{k(d-1)(d-2)\pi^{\frac{d-1}{d-2}}}{2\Lambda\left(2\Gamma\Bigl(\frac{d-1}{2}\Bigr)\right)^{\frac{2}{d-2}}},\label{product02}\\
&\prod_{i=1}^DS_i=\left(\frac{\pi^{\frac{d-1}{2}}}{2\Gamma\Bigl(\frac{d-1}{2}\Bigr)}\right)^{d-1}
                 \left(\frac{M(d-1)(d-2)}{\Lambda}\right)^{d-2}.
\end{align}
The former one has the same behavior with the charged case (\ref{eq:varentropyproduct_of_E-M}). They both only
depend on the cosmological constant $\Lambda$ and horizon topology $k$. One need note that the two
entropy products have different number of horizons $D$, including physical and un-physical (``virtual'') ones.
This type of entropy product relation is also firstly introduced in \cite{Visser:2012wu}. In that paper,
instead of the entropy $S$, horizon area $A$ is used for $S=\frac{1}{4}A$. The latter entropy product, however, is mass dependent, which destroys the well-known mass independence. For
this latter one, the electric charge $Q$ seems to play a switch role: when $Q$ vanishes, the entropy product
turns to mass dependence.

A detailed discussion that considers these two types of entropy products in an asymptotical flat spacetime can be found in the \hyperref[sec:Asymptotical Flat cases]{Appendix}. It is shown that, for the former one, the cosmological constant $\Lambda$ seems to play a switch role: when $\Lambda$ vanishes, the entropy product turns to mass dependence. The phenomenon for additional equalities in thermodynamics also happens in the entropy sum study of multi-horizon black hole \cite{Wang:2013smb}. However, another type of entropy product has the expected electric charge $Q$ dependence.

\section{Higher Dimensional (A)dS Black Holes in $f(R)$ Gravity}
In this section, we will consider the $f(R)$ gravity for a further test of the entropy products of multi-horizon black holes.
\subsection{Charged Solution}
Since the standard Maxwell energy-momentum tensor is not traceless, people failed to derive higher dimensional
black hole/string solutions from $f(R)$ gravity coupled to standard Maxwell field \cite{Xu:2013zpa}. This makes us can only discuss the charged black hole in four dimensions.

Let us consider the action for four dimensional $f(R)$ gravity with Maxwell term,
\begin{align}
  \mathcal{L}=\frac{1}{16\pi G}\int d^4 x \sqrt{-g}[R+f(R)-F_{\mu\nu}F^{\mu\nu}],
\end{align}
where $f(R)$ is an arbitrary function of the scalar curvature $R$. We will focus on the static, spherically
symmetric constant curvature ($R=R_0$) solution with the same metric ansatz as Eq.(\ref{eq:metric ansatz}), in which the horizon function behaves as \cite{Sheykhi:2012zz,Moon:2011hq,Hendi:2011eg,Cembranos:2011sr}
\begin{align}
  V(r)=k-\frac{2\mu}{r}+\frac{q^2}{r^2}\frac{1}{(1+f^{\prime}(R_0))}-\frac{R_0}{12}r^2.
\end{align}
where the cosmological constant of this theory is then have the form $\Lambda_f=\frac{R_0}{4}$. The parameters $\mu$ and $q$ are related to the mass and charge of black hole, respectively. The number of horizons can be $D=4$ at most, including the ``virtual'' horizon. The entropy of all horizons is
\begin{align}
  S_i=\frac{A_i}{4}(1+f^{\prime}(R_0)).
\end{align}
Where $f^{\prime}(R_0)=\left.\frac{\partial f(R)}{\partial R}\right|_{R=R_0}$ and $A_i=4\pi r_i^2$.
We are interested in the entropy product of multi-horizons, which are the roots of the following equation
\begin{align}
\frac{R_0}{12}r^4-kr^2+2\mu r-\frac{q^2}{1+f^{\prime}(R_0)}=0.
\end{align}
Then we can get
\begin{align}
&\sum_{1\leq i<j\leq 4}r_ir_j=-\frac{12k}{R_0},\\
&\prod_{i=1}^4r_i=-\frac{12 q^2}{(1+f^{\prime}(R_0))R_0}.
\end{align}
Obviously these two relationships lead to the following entropy product
\begin{align}
&\sum_{1\leq i<j\leq 4}\sqrt{S_iS_j}=-\frac{12k}{R_0} (1+f^{\prime}(R_0))\pi
                                                              \label{eq:varentropyproduct of 4d charged f(R)}
\intertext{and}
&\prod_{i=1}^4S_i=\frac{144\pi^4q^4(1+f^{\prime}(R_0))^2}{R_0^2}.
\end{align}
After inserting $R_0=4\Lambda_{f}$, the conserved charge dependence of entropy products, $\sum_{1\leq i<j\leq D}(S_{i}S_{j})^{\frac{1}{d-2}}$ and $\prod_{i=1}^DS_i$, are tested in $f(R)$ gravity. They behavior like that in Einstein gravity: the former depends on cosmological constant $\Lambda$ and the horizon topology $k$; while the latter depends on $\Lambda$ and electric charge $Q$. Both of them are independence of mass $M$.

\subsection{Neutral Solution}
Let us consider the Lagrangian of $f(R)$ gravity in higher dimensions
\begin{align}
  \mathcal{L}=\int d^d x \sqrt{-g}(R+f(R)).
\end{align}
After choosing  the same metric ansatz as Eq.(\ref{eq:metric ansatz}), the metric function is \cite{Sheykhi:2012zz}
\begin{align} \label{V(r)22}
  V(r)=k-\frac{2m}{r^{d-3}}-\frac{R_0}{d(d-1)}r^2,
\end{align}
The parameter $m$ is a integral
constant related to the mass (total energy) of the black hole. The cosmological constant is
$\Lambda_{f}=\frac{d-2}{2d}R_0$. The entropy also satisfies the area theorem, and takes the form as
\begin{align} \label{eq:entropy of higher d f(R)}
S_i=\frac{A_i}{4}(1+f^{\prime}(R_0)).\\
\intertext{with the area of every horizon}
A_i=\frac{2\pi^{(d-1)/2}}{\Gamma\left(\frac{d-1}{2}\right)}r_i^{d-2} \notag
\end{align}
where $r_i$ is the horizon and should be the root of Eq.(\ref{V(r)22}), namely,
\begin{equation}
\frac{R_0}{d(d-1)}r^{d-1}-kr^{d-3}+2m=0.
\end{equation}
The black hole at most can possess $D=d-1$ horizons including the ``virtual'' horizon. Then according to Vieta's theorem, we can obtain
\begin{align}
&\sum_{1\leq i<j\leq D}r_ir_j =-\frac{k\,d(d-1)}{R_0}\\
&\prod_i^Dr_i =(-1)^D\frac{2m\,d(d-1)}{R_0}
\end{align}
Followig the same procedure we inverse (\ref{eq:entropy of higher d f(R)}) as
\begin{equation}
r_i=\left(\frac{2\Gamma\Bigl(\frac{d-1}{2}\Bigr)S_i}{\pi^{\frac{d-1}{2}}(1+f^{\prime}(R_0))}\right)^{\frac{1}{d-2}}.
\end{equation}
Then we get this type of entropy product
\begin{align} \label{eq:varentropyproduct_of_f(R)}
\sum_{1\leq i<j\leq D}(S_{i}S_{j})^{\frac{1}{d-2}}=
     -\frac{kd(d-1)}{R_0}
     \left(\frac{\pi^{\frac{d-1}{2}}(1+f^{\prime}(R_0))}{2\Gamma\Bigl(\frac{d-1}{2}\Bigr)}\right)^{\frac{2}{d-2}}
\end{align}
We note that $R_0=\frac{2d\Lambda_{f}}{d-2}$, which means the above one only depends on the cosmological constant. To consider (\ref{eq:varentropyproduct_of_E-M}), (\ref{product02}), (\ref{eq:varentropyproduct of 4d charged f(R)}) and (\ref{eq:varentropyproduct_of_f(R)}) together, one can treat them as one type of entropy product, consisting of the sum of “part” of the entropy product and only having different numbers of
the possible horizons $D$. Furthermore, this type of entropy product, $\sum_{1\leq i<j\leq D}(S_{i}S_{j})^{\frac{1}{d-2}}$  is always conserved quantities independent. That means this quantity does not relate to the mass or electric charge distribution. This entropy product only depends on the cosmological constant and horizon topology in both Einstein gravity and $f(R)$ gravity with or without Maxwell field source in the spacetime. In this sense, we conclude the entropy product, $\sum_{1\leq i<j\leq D}(S_{i}S_{j})^{\frac{1}{d-2}}$, is ``universal''.

However, the entropy product
\begin{align}
\prod_{i=1}^DS_i=\left(\frac{\pi^{(d-1)/2}(1+f^{\prime}(R_0))}{2\Gamma\left(\frac{d-1}{2}\right)}\right)^{d-1}
                 \left(\frac{2m\,d(d-1)}{R_0}\right)^{d-2}
\end{align}
is mass dependent, even including the effect of the un-physical ``virtual'' horizon. It is quite different from the suggestion provided in \cite{Cvetic:2010mn} and \cite{Visser:2012wu}, where the mass independence of entropy product $\prod_{i=1}^DS_i$ always fails for the uncharged black holes.

\section{Conclusion}
We have discussed the entropy products of the higher dimensional (A)dS black holes in Einstein-Maxwell and $f(R)$(-Maxwell) gravity, which have
possessed similar formula and can be concluded as below
\begin{enumerate}
  \item There exists a ``universal'' quantity $\sum_{1\leq i<j\leq
      D}(S_{i}S_{j})^{\frac{1}{d-2}}$, which only depends on the cosmological constant $\Lambda$ and background
      topology $k$ and does not depend on the conserved charges $Q$, nor even the mass $M$.
  \item Being different from \cite{Cvetic:2010mn} and \cite{Visser:2012wu}, another entropy product $\prod_{i=1}^DS_i$ in our case has an unexpected behavior. It is shown that the electric charge $Q$ plays an important role in this entropy product. The entropy product of charged black holes only depend on the electric charge $Q$ and it is mass independence. When $Q$ vanishes in the solution, it becomes mass dependent, even when including the effect of the unphysical virtual horizons. In this sense, the ``universal property'' of this entropy product is destroyed.
\end{enumerate}

After having a whole look at entropy product or entropy sum to multi-horizons of the black hole, one can find that they all only depend on the conserved charges: the electric charge $Q$, the angular momentum $J$, or the cosmological constant $\Lambda$ (which can be treated as pressure in AdS spacetime after interpreting the mass of the black hole as  the thermodynamic enthalpy rather than internal energy of the system). The use of the entropy relation now is not clear, but it may be relevant to the holographic
description of black holes in the future. A further test on the entropy product of the rotating black hole and its reduced static black hole is planned to be a future task.

\begin{acknowledgments}
This work is partially supported by the Natural Science Foundation of China (NSFC) under Grant No.11075078 and
by the project of knowledge innovation program of Chinese Academy of Sciences.
\end{acknowledgments}

\appendix
\section{Higher Dimensional Einstein-Maxwell Black Holes in Asymptotical Flat Spacetime}
\label{sec:Asymptotical Flat cases}
We should note that when we ``switch off'' the cosmological constant $\Lambda$ in higher dimensional (A)dS
Einstein-Maxwell solutions, the result is the generalized ordinary Reissner-Nordst{\"o}rm black hole. The
Lagrangian is
\begin{align}
\mathcal{L}=\frac{1}{16\pi G}\int d^d x \sqrt{-g}[R-F_{\mu\nu}F^{\mu\nu}]
\end{align}
Similarly we take the same metric ansatz (\ref{eq:metric ansatz}) and the metric function
\begin{align}
V(r)=k-\frac{2M}{r^{d-3}}+\frac{Q^2}{r^{2(d-3)}}.
\end{align}
Immediately we get
\begin{align}
&\prod_{i=1}^{2(d-3)}r_i=\frac{Q^2}{k},\\
&\sum_{1\leq i_1< i_2\cdots< i_a\leq D}
 \,\prod_{a=1}^{d-3}r_{i_a}=(-1)^{d-3}\,\frac{-2M}{k},
\end{align}
where we have considered that the number of the horizon is maximum. As we know, when $k=0$, i.e. the black hole horizon with Ricci flat topology, $\int d\Omega^{2}_{d-2}$ can be arbitrary positive \cite{Brill:1997mf}, which makes the entropy arbitrary positive as well. However, this is out of our discussion. Thus, in the following discussion we will consider with $k=\pm 1$, which correspond to spherical or hyperbolic horizons. The above two equations directly derive the entropy product
\begin{align}
  \sum_{1\leq i_1< i_2\cdots< i_a\leq D}
  \left(\prod_{a=1}^{d-3}S_{i_a}\right)^{\frac{1}{d-2}}=\left(\frac{\pi^{\frac{d-1}{2}}}{2\Gamma\Bigl(\frac{d-1}{2}\Bigr)}\right)^{\frac{d-3}{d-2}}(-1)^{d-2}\frac{2M}{k}.
\end{align}
and
\begin{align}
  \prod_{i=1}^{2(d-3)}S_i=\left(\frac{\pi^{(d-1)/2}}{2\Gamma\left(\frac{d-1}{2}\right)}\right)^{2(d-3)}
                          \left(\frac{Q^2}{k}\right)^{d-2}.
\end{align}
Comparing with the (A)dS case, we find the type of entropy product,\\
$\sum_{1\leq i_1< i_2\cdots< i_a\leq D}\left(\prod_{a=1}^{d-3}S_{i_a}\right)^{\frac{1}{d-2}}$, becomes mass $M$ and horizon topology $k$ dependence and does not depend on the cosmological constant $\Lambda$. This entropy product can be treated as the same type with (\ref{eq:varentropyproduct_of_E-M}), both of which consist of the sum of the ``part'' entropy product. Thus, for this entropy product, the cosmological constant $\Lambda$ seems to play a reversed role instead of $Q$: when $\Lambda$ vanishes, the entropy product becomes mass dependent. In other word, mass maybe be the next order of $\Lambda$ for this entropy product. The is not a new phenomenon for additional equalities in thermodynamics, as it also happens in
the entropy sum study of multi-horizon black hole \cite{Wang:2013smb}. Another type of charged black hole $\prod_{i=1}^{2(d-3)}S_i$ has the same electric charge $Q$ dependence with the (A)dS case.

Let us continue to consider the case that even the electric charge $Q=0$ i.e. the ``pure'' gravity field. In fact it is a higher dimensional Schwarzschild black hole with the metric function
\begin{align}
V(r)=k-\frac{2M}{r^{d-3}}.
\end{align}
We have one ``physical'' horizon (event horizon) and $(d-4)$ ``virtual'' horizons. There is only one type of entropy product as
\begin{align}
      \prod_{i=1}^{d-3}S_i=\left(\frac{\pi^{(d-1)/2}}{2\Gamma\left(\frac{d-1}{2}\right)}\right)^{d-3}
                          \left(\frac{-2M}{k}\right)^{d-2}.
\end{align}
Obviously, the entropy product is mass dependence, which is consistent with the case of uncharged (A)dS black hole.

\providecommand{\href}[2]{#2}\begingroup
\footnotesize\itemsep=0pt
\providecommand{\eprint}[2][]{\href{http://arxiv.org/abs/#2}{arXiv:#2}}

\end{document}